\documentclass[sigconf]{acmart}
\renewcommand\footnotetextcopyrightpermission[1]{} 
\usepackage{algorithm}
\usepackage{multirow}
\usepackage{subcaption}
\usepackage[absolute,overlay]{textpos}

\usepackage[noend]{algorithmic}
\usepackage{amsmath}

\AtBeginDocument{%
	\providecommand\BibTeX{{%
			\normalfont B\kern-0.5em{\scshape i\kern-0.25em b}\kern-0.8em\TeX}}}

\setcopyright{acmcopyright}
\copyrightyear{2018}
\acmYear{2018}
\acmDOI{XXXXXXX.XXXXXXX}

\acmConference[ICPP'2022]{Make sure to enter the correct
	conference title from your rights confirmation emai}{August,
	2022}{Bordeaux,France}
\acmPrice{15.00}
\acmISBN{978-1-4503-XXXX-X/18/06}

\begin{document}
\begin{textblock*}{10cm}(8cm,26cm) 
   {\huge To appear in ICPP'2022}
\end{textblock*}
	\begin{sloppypar}
		
		\title{FedHiSyn: A Hierarchical Synchronous Federated Learning Framework for Resource and Data Heterogeneity}
		
		\author{Guanghao Li$^{1}$, Yue Hu$^{1}$, Miao Zhang$^1$, Ji Liu$^2$\footnotemark[1], Quanjun Yin$^1$, Yong Peng$^1$\footnotemark[1], Dejing Dou$^2$}
        \affiliation{
        \textsuperscript{\rm 1}National University of Defense Technology, 
        \textsuperscript{\rm 2}Baidu Inc., China\\
        * Corresponding author (Ji Liu: liuji04@baidu.com and Yong Peng: yongpeng@nudt.edu.cn)
        \country{}
        }
		

		\renewcommand{\shortauthors}{Li et al.}
		
		\begin{abstract}
			Federated Learning (FL) enables training a global model without sharing the {decentralized} raw data stored on {multiple} devices to protect data privacy. Due to the 
			{diverse capacity of the} devices,
			FL frameworks struggle to tackle the problems of straggler effects and outdated models
			. In addition, the data heterogeneity incurs severe accuracy degradation of the global model {in the FL training process}. To address {aforementioned} issues, we propose a hierarchical synchronous FL framework, i.e., FedHiSyn. FedHiSyn first clusters all available devices into a small number of categories based on their computing capacity. After a certain interval of local training, the models trained in different categories are simultaneously uploaded to a central server. Within a single category, the devices communicate the local updated model weights to each other based on a ring topology. 
			As the efficiency of training in the ring topology prefers devices with homogeneous resources, the classification based on the computing capacity mitigates the impact of straggler effects. 
			{Besides, the combination of the synchronous update of multiple categories and the device communication within a single category help address the data heterogeneity issue while achieving high accuracy.}
			We evaluate the proposed framework based on MNIST, EMNIST, CIFAR10 and CIFAR100 datasets and diverse heterogeneous settings of devices. Experimental results show that FedHiSyn outperforms six baseline methods, {e.g.,} FedAvg, SCAFFOLD, and FedAT, in terms of training accuracy and efficiency.  
		\end{abstract}
		
%
		
		

		\maketitle
		
		\section{Introduction}
		
		As the number of smartphones and  smart wearable devices has grown rapidly in the past few years, {large amount of} data has been collected and stored on those edge devices with storage and data processing functions. As an emerging learning paradigm, {F}ederated {L}earning (FL) can realize collaborative machine
		learning by making full use of computation, storage and data resources of edge devices while {protecting} data privacy ~\cite{DBLP:conf/aistats/McMahanMRHA17,Liu2022From}. FL has been used in a number of application areas, including predicting human activities ~\cite{DBLP:conf/bigdataconf/ChenNSR20,DBLP:journals/tnn/ChenSJ20}, learning sentiment ~\cite{DBLP:conf/nips/SmithCST17}, language processing ~\cite{DBLP:journals/corr/abs-1811-03604,DBLP:conf/iclr/LiSBS20} and enterprise infrastructures ~\cite{DBLP:journals/corr/abs-2007-10987}. 
		
		\begin{figure}[h]
			\centering
			\includegraphics[width=\linewidth]{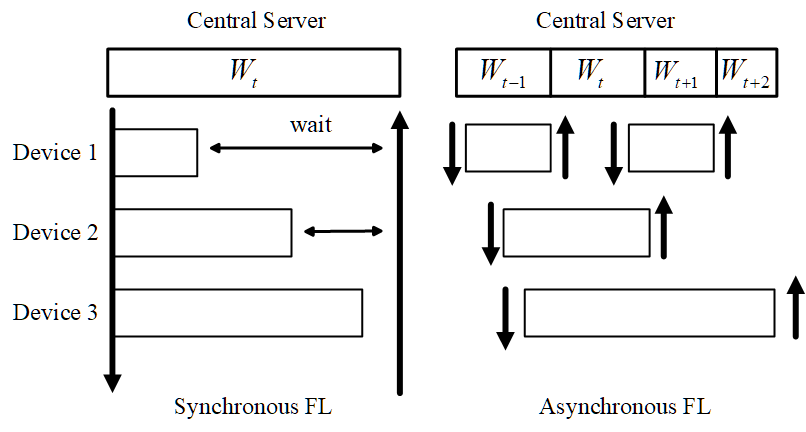}
			\vspace{-4mm}
			\caption{The training procedures of synchronous FL and asynchronous FL}
			\vspace{-4mm}
			\label{fl}
		\end{figure}
		
		In a typical FL framework, the participating devices first conduct local training based on their own data and then upload the weight updates to a central server. The central server aggregates the models  from the various devices in each round and then broadcasts updated model parameters to the devices. During the entire training process of the {FL}, the training data on each device is kept locally and not transmitted, which protects data privacy ~\cite{DBLP:conf/icc/HaoLXLY19}. 
		
		FL often involves a large number of devices, which usually feature highly 	heterogeneous hardware resources (CPU, memory, and network resources) and Non-IID ({N}on {I}ndependent-{I}dentically-{D}istributed) data. Existing FL frameworks can be {classified} into synchronous FL (e.g., FedAvg ~\cite{DBLP:conf/aistats/McMahanMRHA17}) and asynchronous FL (e.g., FedAsync ~\cite{DBLP:conf/trustcom/LiZCWJ21}). The training process of the two frameworks is shown in Figure~\ref{fl} where $\mathcal{W}_t$ denotes the $t$-th update of the weights $\mathcal{W}$ of the global model. Synchronous FL tends to result in the straggler effect~\cite{liu2022Efficient} (especially on the scale of hundreds of {heterogeneous} devices), causing the device to go into an idle state. Asynchronous FL avoids devices falling into the idle state, but the more frequent communication between comparatively more powerful devices and the server can cause the crash of the server~\cite{DBLP:conf/icc/ShiZN20}. The outdated models transmitted from the less powerful devices to the server also affect the training process of the global model~\cite{DBLP:conf/trustcom/LiZCWJ21}. And beyond that,  in both frameworks, the Non-IID data stored on the devices can cause significant differences of weights updated by {the} devices, which greatly affects the training accuracy of the final model~\cite{DBLP:journals/corr/abs-1806-00582}.
		
		In order to overcome the above defects caused by heterogeneous resources and Non-IID data, we design and implement FedHiSyn,  a two-layer synchronous FL framework. FedHiSyn first clusters devices based on the computing power of each device. Then,  devices within  single a category communicate their locally trained models to others following a ring topology. Specifically, each device transfers the trained model to the next device in the ring and the latter device continues training the model based its own data. Finally, all categories simultaneously send model updates to the server after a certain interval. {FedHiSyn utilizes both device-to-device communication and device-server communication  and therefore it is a combination of the centralized  and  decentralized FL framework.}
		
		Our {major design principle} is to use communication between devices in exchange for communication between devices and the server, thereby increasing training speed and reducing communication costs. We {show} that the training mode based on the ring topology 
		{corresponds to high accuracy,}
		and reduces {the number of model training rounds}. However, the  training efficiency {of FL} is {significantly} affected by {the heterogeneous resources of devices}. When there is a large difference in computing power between devices, devices with strong computing power need to wait for a long time before sending information to the next node. For this reason, FedHiSyn {clusters devices} with similar computing capacity {into} a same group {to avoid} the communication {among devices of highly heterogeneous resources}. At the end of each training round in our approach, all groups perform a synchronous update strategy. Experimental results show that our algorithm is robust to large-scale scenarios and highly heterogeneous devices.

		We make the following {major} contributions in this paper:
		\begin{itemize}
			\item We design and implement a new tiered FL framework, named FedHiSyn, which { combines  the centralized FL framework and the decentralized FL framework. On the top layer, this paper tries to apply the device clustering based computing capacity to circumvent the problem of resource heterogeneity. Within single groups, it utilizes the communication among devices to reduce the communication load of the server and mitigate the impact of Non-IID data}. 
			
			\item We provide a rigorous theoretical analysis of our proposed methods for strong-convex objectives. Our analysis shows that FedHiSyn has a provable convergence guarantee {and can converge to the optimal solution faster than FedAvg}. 
			\item We evaluate FedHiSyn extensively on four datasets (including MNIST, EMNIST, CIFAR-10 and CIFAR100) and the experimental results show that FedHiSyn has better test accuracy  and lower communication cost,  compared with six baselines such as Fedprox and FedAT.  For instance, 			FedHiSyn improves the test accuracy by up to 10.28\%, reducing the communication cost by up to 7.7$\times$ compared with FedAvg. 
		\end{itemize}
		
		\section{Related Work}
		
		In this section, we introduce works about FL frameworks on Non-IID data and resource heterogeneity. 
		
		\subsection{Non-IID Data}
		
		McMahan et al.~\cite{DBLP:conf/aistats/McMahanMRHA17} first proposed the basic framework of FL and designed a algorithm called FedAvg. The authors found that Non-IID data had an impact on the prediction accuracy of the global model.  Zhao et al. ~\cite{DBLP:journals/corr/abs-1806-00582} showed that the accuracy of FedAvg decreases significantly with the increase of data heterogeneity.  In order to cope with the challenges posed by Non-IID data, FedDyn~\cite{DBLP:conf/iclr/AcarZNMWS21} and SCAFFOLD ~\cite{DBLP:conf/icml/KarimireddyKMRS20} use regularization methods to estimate the global knowledge of data distributions of all devices, respectively. However, this approach has a large bias when only a small number of devices participate in every training round. Momentum \cite{JinAccelerated2022} can be exploited to improve the accuracy of FL, which can be combined with other methods \cite{DBLP:conf/icml/KarimireddyKMRS20}. CMFL ~\cite{DBLP:conf/icdcs/WangWL19}, Oort~\cite{DBLP:conf/osdi/LaiZMC21} and Favor ~\cite{DBLP:conf/infocom/WangKNL20} neutralize the influence of Non-IID and speed up convergence by selecting a group of ``excellent" devices to participate in each round of training. However, these methods tend to ignore the precious data stored on a few devices. FedRep ~\cite{DBLP:conf/icml/CollinsHMS21} and FedMD ~\cite{DBLP:journals/corr/abs-1910-03581} make use of transfer learning and knowledge distillation, respectively, to establish different models for each device, making it difficult for new devices to select an appropriate model for initialization. 
		
		\subsection{Heterogeneous resources}
		
		The original assumption of FL was that the computing resources of all devices were the same, but resource heterogeneity usually exists in a real federated environment. Resource heterogeneity can cause the straggler problems in traditional synchronous FL frameworks ~\cite{DBLP:conf/opml/ChaiFFAZBLC19}. FedCS ~\cite{DBLP:conf/icc/NishioY19} and FedMCCS ~\cite{DBLP:journals/iotj/RahmanTMT21} choose devices with sufficient computing resources to prevent devices from falling into idle waiting and wasting computing resources. However, strict limits on customer choice can lead to fewer participants, slowing down the rate of convergence. FedProx ~\cite{DBLP:journals/iotj/RahmanTMT21} uses partial local update work of stragglers to alleviate model degradation. However, choosing a perfect number of local epochs for each device is challenging in real-world applications. Asynchronous FL has an inherent advantage over synchronous FL in addressing drop-off effects because models can be aggregated without waiting for stragglers. For example, Aso-fed ~\cite{DBLP:conf/bigdataconf/ChenNSR20} updates the global model asynchronously to deal with stragglers. However, gradient obsolescence is not taken into account, which may threaten the convergence of the model.  FedAsync~\cite{DBLP:journals/corr/abs-1903-03934} combines a function of staleness with asynchronous update protocol. However, the devices still need to transmit a large amount of data to the server, which may cause the server to crash.  FedAT ~\cite{DBLP:conf/sc/ChaiC00CR21} combines synchronous training and asynchronous training to resolve straggler problems and alleviate communication bottleneck.  It is a compromise solution and cannot fundamentally solve the problem. 
		
		At present, there are few studies that consider both the challenges of device heterogeneity and Non-IID data. CSAFL ~\cite{DBLP:conf/ijcnn/ZhangDLLRCTW21} divides devices with similar data distribution into  groups for asynchronous communication according to gradient information.  But the reliability of grouping basis is difficult to guarantee. WKAFL \cite{DBLP:journals/corr/abs-2203-01214} estimates global information to solve the Non-IID problem before asynchronous communication, but the effect is highly related to the estimation accuracy of global information. FedDUAP \cite{Zhang2022FedDUAP} exploits the data on the server with the consideration of the Non-IID data on devices to improve the accuracy while it cannot well address the device heterogeneity.
		
		{The above work only utilizes the communication between the server and a large number of devices, and does not utilize idle channels between devices. And a fully decentralized FL that only utilizes intercommunication between devices does not have a training control center. It is susceptible to the topology between devices leading to a loss of part of the training speed and model accuracy~\cite{DBLP:conf/uic/MalladiLSSH21}.}
		
		Our proposed FedHiSyn has a two-tiered architecture {combinates  the centralized FL framework and the decentralized FL framework}. In the first layer, devices are divided into different groups based on their computing capabilities to avoid the degration of training efficiency caused by the heterogeneity of device resources. At the second tier, devices whthin each group communicate with each other based on a ring topology to mitigate Non-IID issues. At intervals, all devices simultaneously send updated models to the server. Additionally, FedHiSyn is robust to large numbers of devices, partial participation, and imbalanced data. 
		\section{Background and Motivation}
		In this section, we first introduce the traditional framework of FL, i.e., FedAvg, and then discuss the key observations that motivate the design of FedHiSyn.  
		\subsection{FedAvg Framework}
		FL usually consists of two main components: a central server and $C$  devices. In each round, only part of all devices, denoted as a set $\mathcal{S}$, are selected for each round of training. The whole training process of FedAvg can be divided into the following steps:
		
		(1) The central server broadcasts the parameters $\mathcal{W}_G$ of the global model to be trained on the participating devices.
		
		(2) Each participating device $i$ updates the parameter $ \mathcal{W}_i $ of the local model using its local data by minimizing the empirical risk represented by Eq. (1):
		
		\begin{equation}
		F(\mathcal{W}) = \sum_{i=1 }^{C} p_i F_i(\mathcal{W}),
		\end{equation}  
		where $F_i(.)$ represents the local loss function based on training set of device $i$, and $p_i$ is the weight of the $i$-th device such that $p_i  \ge  0$ and $\sum_{i=1}^C p_i = 1$. The update formula for $\mathcal{W}_i$ is represented as Eq. (2):
		\begin{equation}
		\mathcal{W}_i^{r+1} = \mathcal{W}_G^{r} - \eta\nabla F_i(\mathcal{W}_G^{r}),
		\end{equation} 
		where $\mathcal{W}_i^{r+1}$ is the model in Round  $r+1$, and $\eta$ is the learning rate.
		
		(3) After completing the local training, each participating device $i$ in Round $r$ uploads the accumulated parameters $ \mathcal{W}_i^{r+1} $ to the central server. Then, the central server aggregates the received parameters and updates the global model according to Eq. (3):
		\begin{equation}
		\mathcal{W}_{{G}}^{r+1} = \sum_{i=1}^{\left\|\mathcal{S}\right\|}\frac{n_i}{N} \mathcal{W}_i^{r+1},
		\end{equation} 
		where $\left\|\mathcal{S}\right\|$  is the number of participating devices, $N$ is the number of the total training samples across all devices in this round, and $n_i$ represents the size of the local dataset on Device $i$.
		%
		%
		%
		%
		%
		%
		%
		
		The aforementioned three steps iterate till the shared model converges. We can find that the raw training data are always kept on the participating devices during the overall training process. Thus, data privacy is well preserved.
		\subsection{Motivation}
		In a typical real-world scenario, the data stored on different devices does not follow an IID distribution. Although FedAvg can work with partial device participation at each training round, training on Non-IID data may diverge each device towards its optimal local model as opposed to achieving an optimal global one. 
		
		\begin{figure}[htbp]
			\begin{minipage}[t]{0.5\linewidth}
				\centering
				\includegraphics[width=\textwidth]{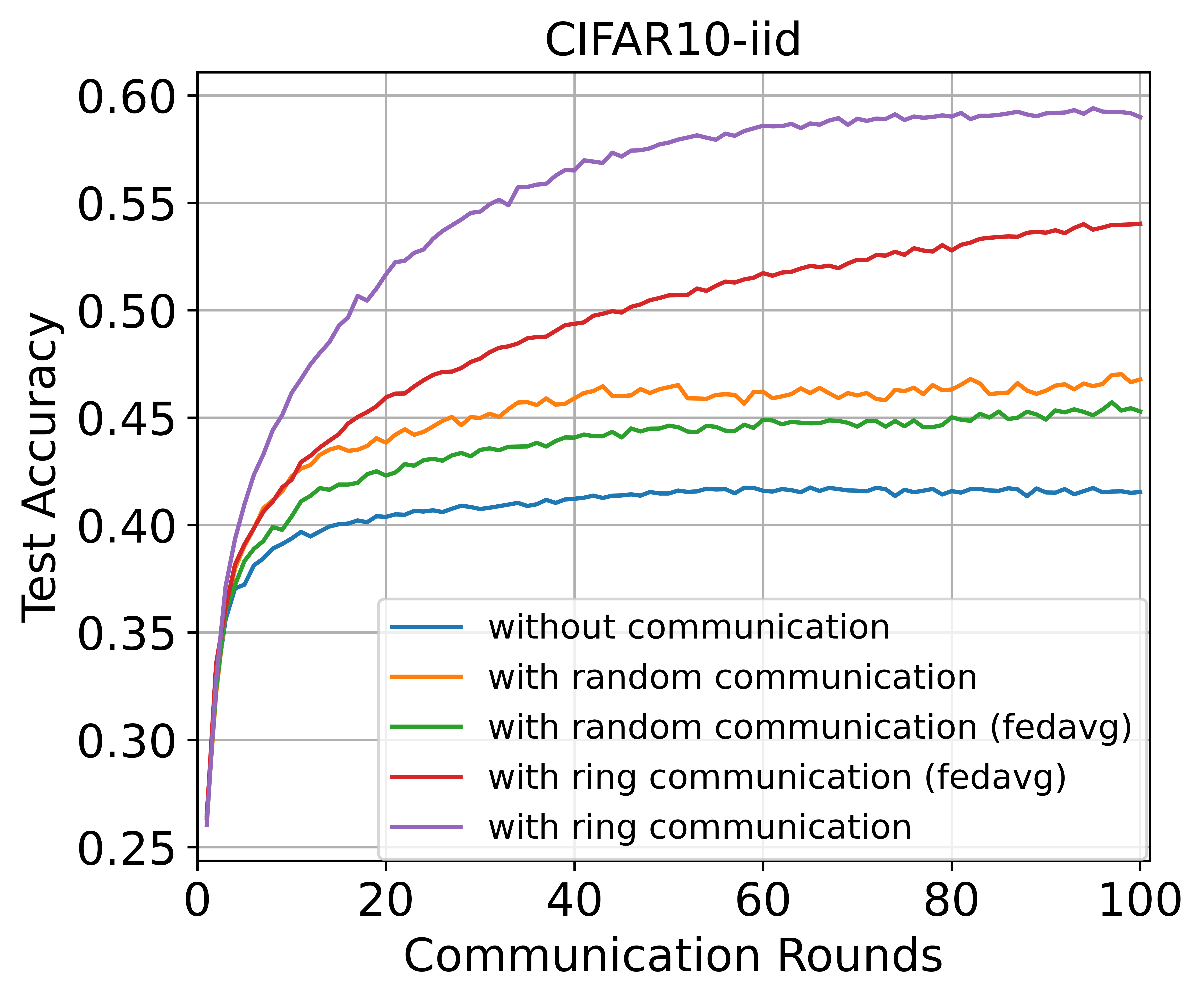}
				\centerline{(a) CIFAR10-IID}
			\end{minipage}%
			\begin{minipage}[t]{0.5\linewidth}
				\centering
				\includegraphics[width=\textwidth]{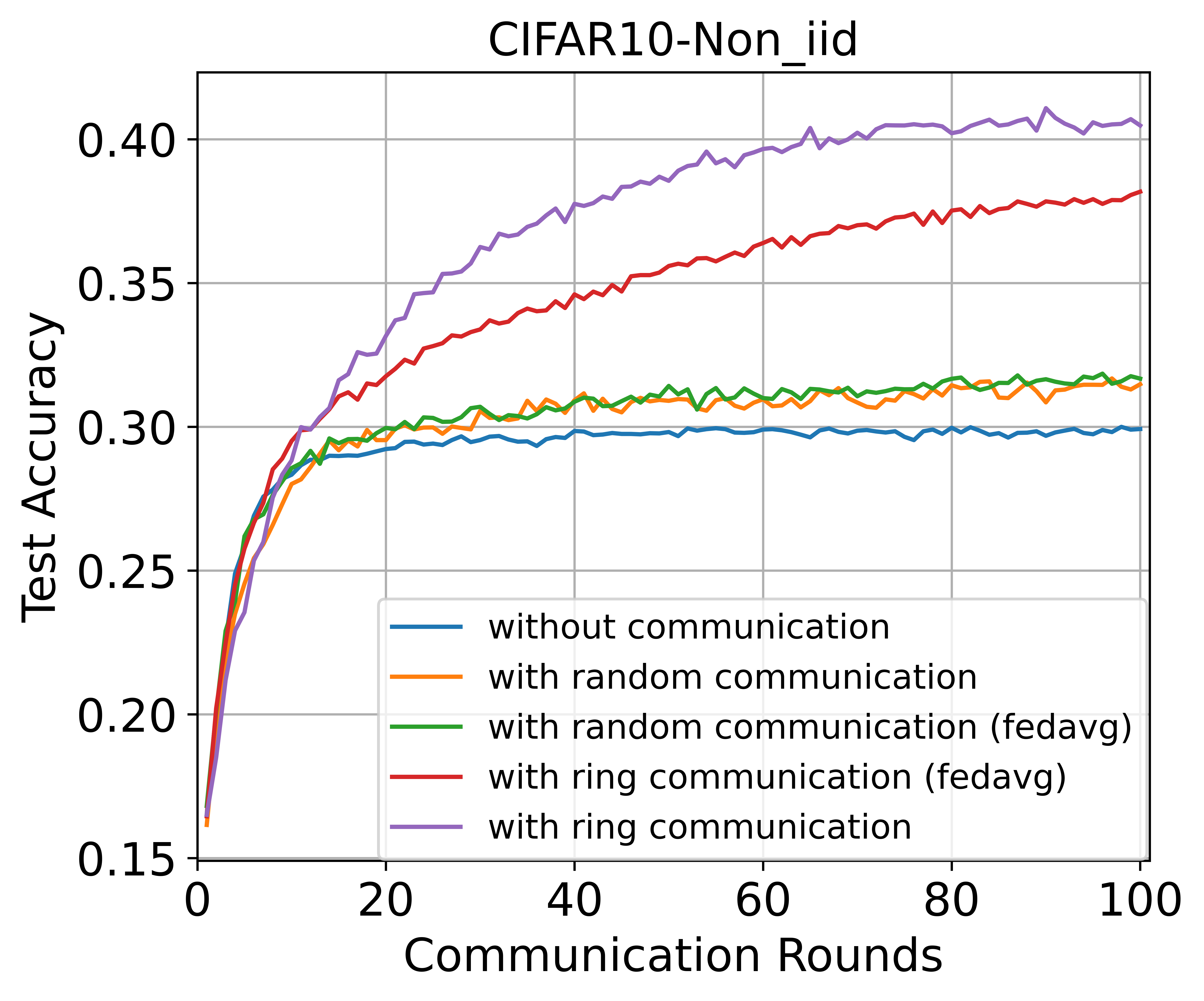}
				\centerline{(b) CIFAR10-Non-IID}
			\end{minipage}
			\vspace{-4mm}
			\caption{Training accuracy in different cases of device communication.}
			\vspace{-9mm}
			\label{mt1}
		\end{figure}
		
		With the same initialization parameters, the divergence in Round $r$ between the parameters $\mathcal{W}_c^r$ of centralized training with all data gathered together and the parameters $\mathcal{W}_f^r$ of traditional FedAvg training mainly comes from Eq. (4):
		\begin{equation}
		D = \sum_{i=1}^{C}\sum_{j=1}^{n_j}\left\|p^i(y=j)-p(y=j)\right\|,
		\end{equation} where $n_j$ is the number of the data classes in each devices $i$, $p^i(y=j)$ is the probability distribution of Label $j$ on Device $i$, and $p(y=j)$ is the probability distribution of Label $j$ on the overall dataset~\cite{DBLP:journals/corr/abs-1806-00582}. The larger the gap between the data distribution on the devices and the overall data distribution is, the lower the accuracy of the final model will be.

		Since data on devices is not available, the indicator $D$ in Eq. (4) cannot be directly calculated. Instead, we exploit another empirical method in this paper. We assume that the data distributions of the training set and test set of overall data are the same. The accuracy of a fully trained model on a device in the  overall test set to some extent represents the difference between the label distribution on the device and the overall label distribution. Therefore, the higher the accuracy of the model trained on a device is, the closer the data label distribution of the device is to the overall data distribution, which we deem that $D$ should be smaller. 
		
		\textbf{Observation 1. } \textit{In the context of FL, the model trained through communication between devices will be more accurate than the model trained on individual devices separately. } 
		
		%
		%
		%

		 To validate Observation 1, we conduct some experiments to identify the influence of communication between devices on the model accuracies. {The resources of devices are set to be homogeneous in this set of experiments.} We consider both IID and Non-IID data settings across devices, and the test data is identically distributed as all training data. We consider comparative five cases: no communication between devices, random communication between devices and then averaging, random communication, ring-topology based communication between devices and averaging, and ring-topology based communication. In our setting, the device will directly use the model received from the last device for local training. Averaging means that the model received from another device  will be aggregated with the local model and then trained locally. We take the average accuracy of models on 100 devices after 50 rounds of training as an estimate of $D$. As shown in Figure 2, in both IID and Non-IID settings, the accuracy of models trained through communication between devices based on the ring topology is significantly higher than those on random communication or no communication. Different parameter aggregation methods will also affect the final training accuracy. Using the received model directly for local training has higher training accuracy than aggregating the received model, especially in the setting of  communication based on the ring topology. This observation leads us to the motivation that we could exploit the communication between devices to make the model more informative so as to decrease $D$ and alleviate the Non-IID problem. 
		
		
		\begin{figure}[htbp]
			\begin{minipage}[t]{0.5\linewidth}
				\centering
				\includegraphics[width=\textwidth]{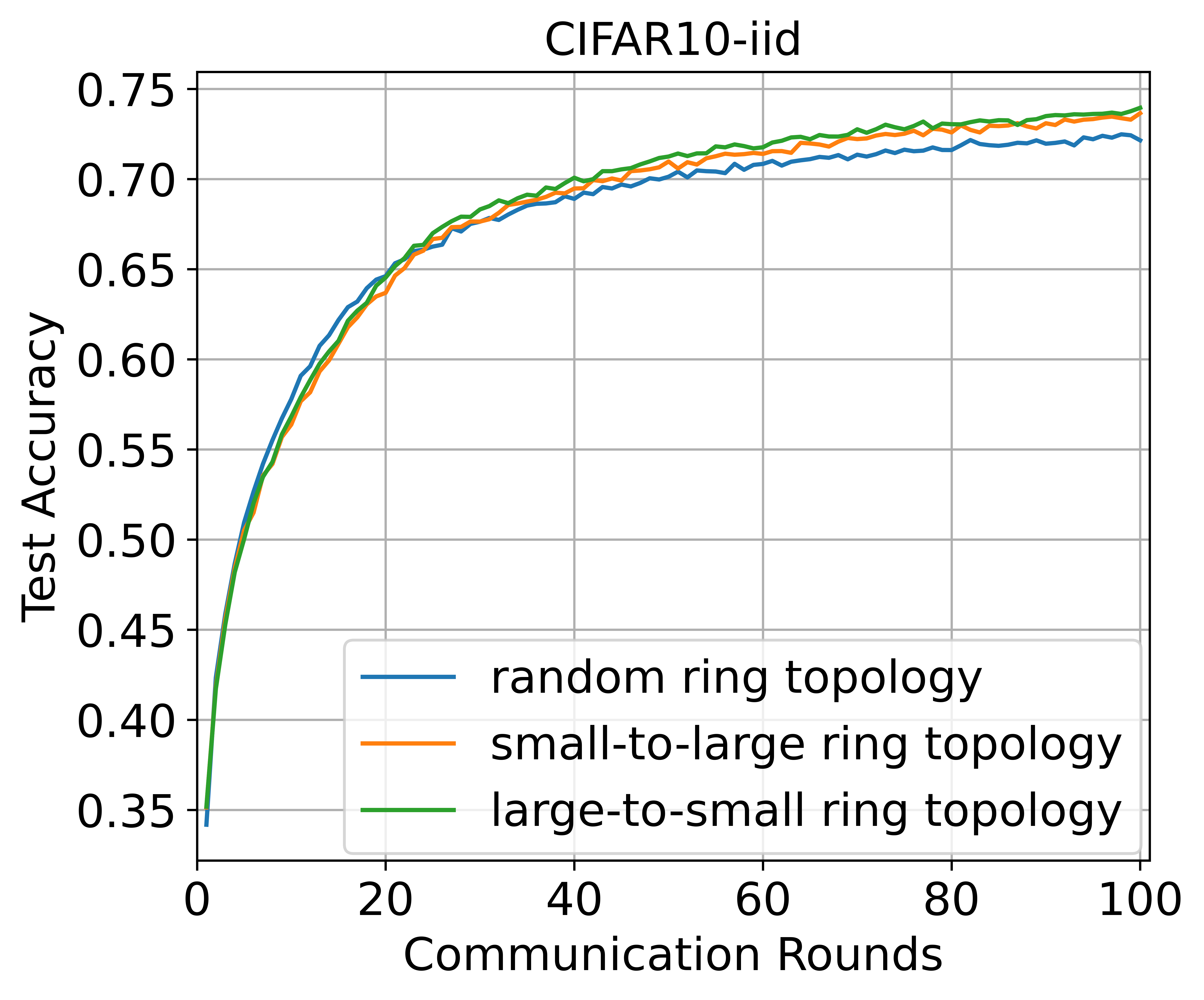}
				\centerline{(a) CIFAR10-IID}
			\end{minipage}%
			\begin{minipage}[t]{0.5\linewidth}
				\centering
				\includegraphics[width=\textwidth]{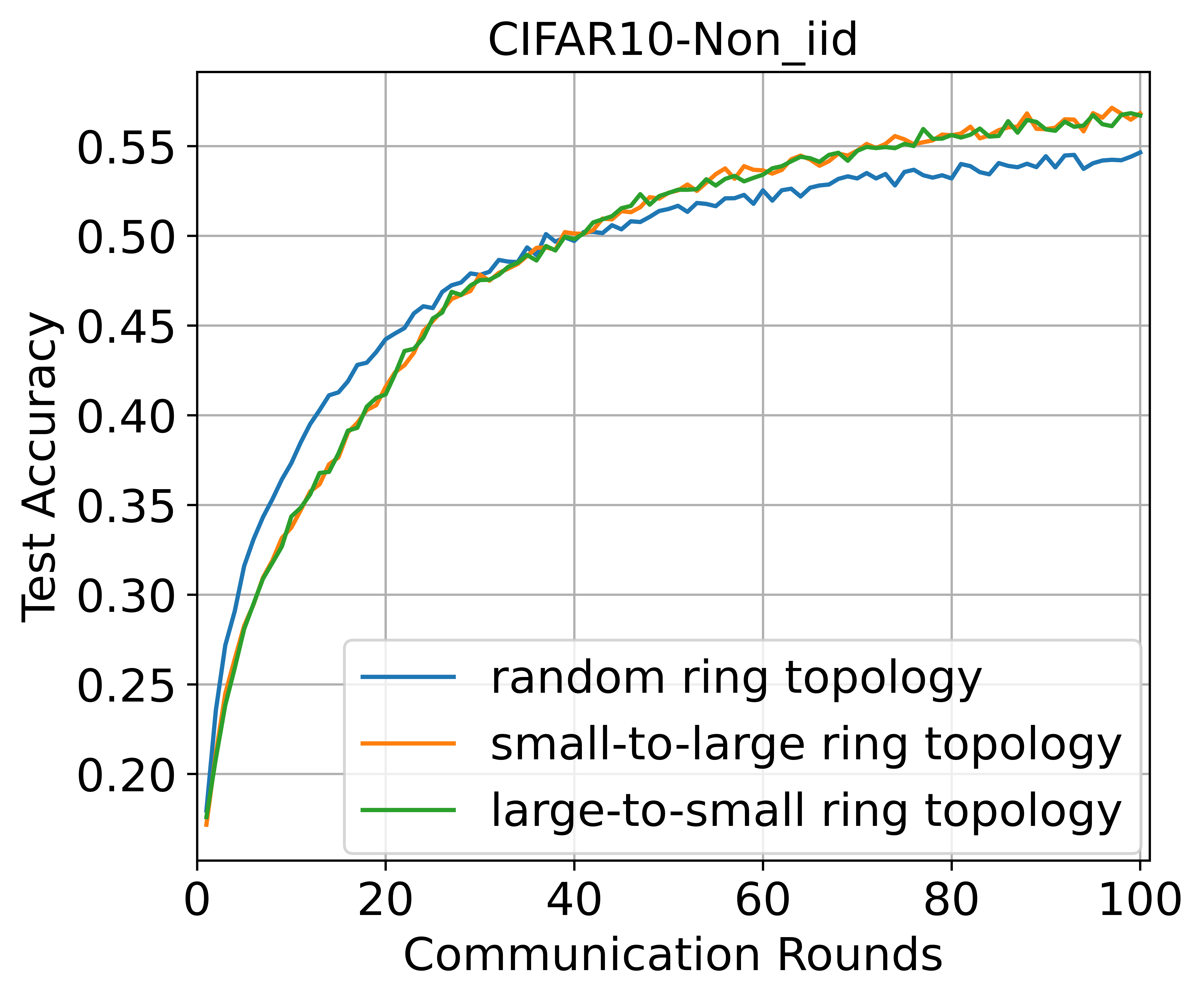}
				\centerline{(b) CIFAR10-Non-IID}
			\end{minipage}
			\vspace{-3mm}
			\caption{The impact of different topological organizations on the training model accuracy.}
			\vspace{-5mm}
			\label{mt3}
		\end{figure}
		
		\begin{figure}[htbp]
			\begin{minipage}[t]{0.5\linewidth}
				\centering
				\includegraphics[width=\textwidth]{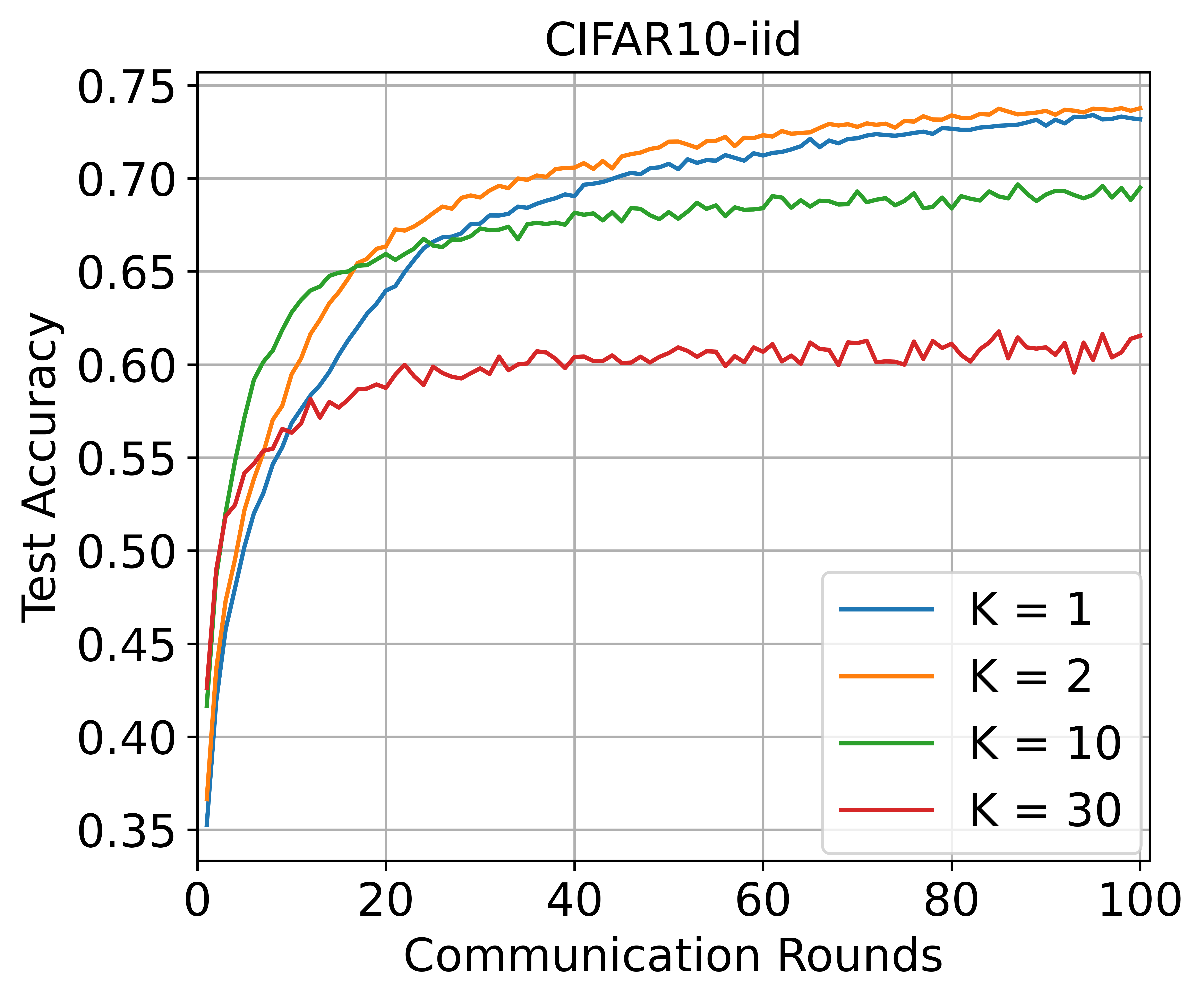}
				\centerline{(a) CIFAR10-IID}
			\end{minipage}%
			\begin{minipage}[t]{0.5\linewidth}
				\centering
				\includegraphics[width=\textwidth]{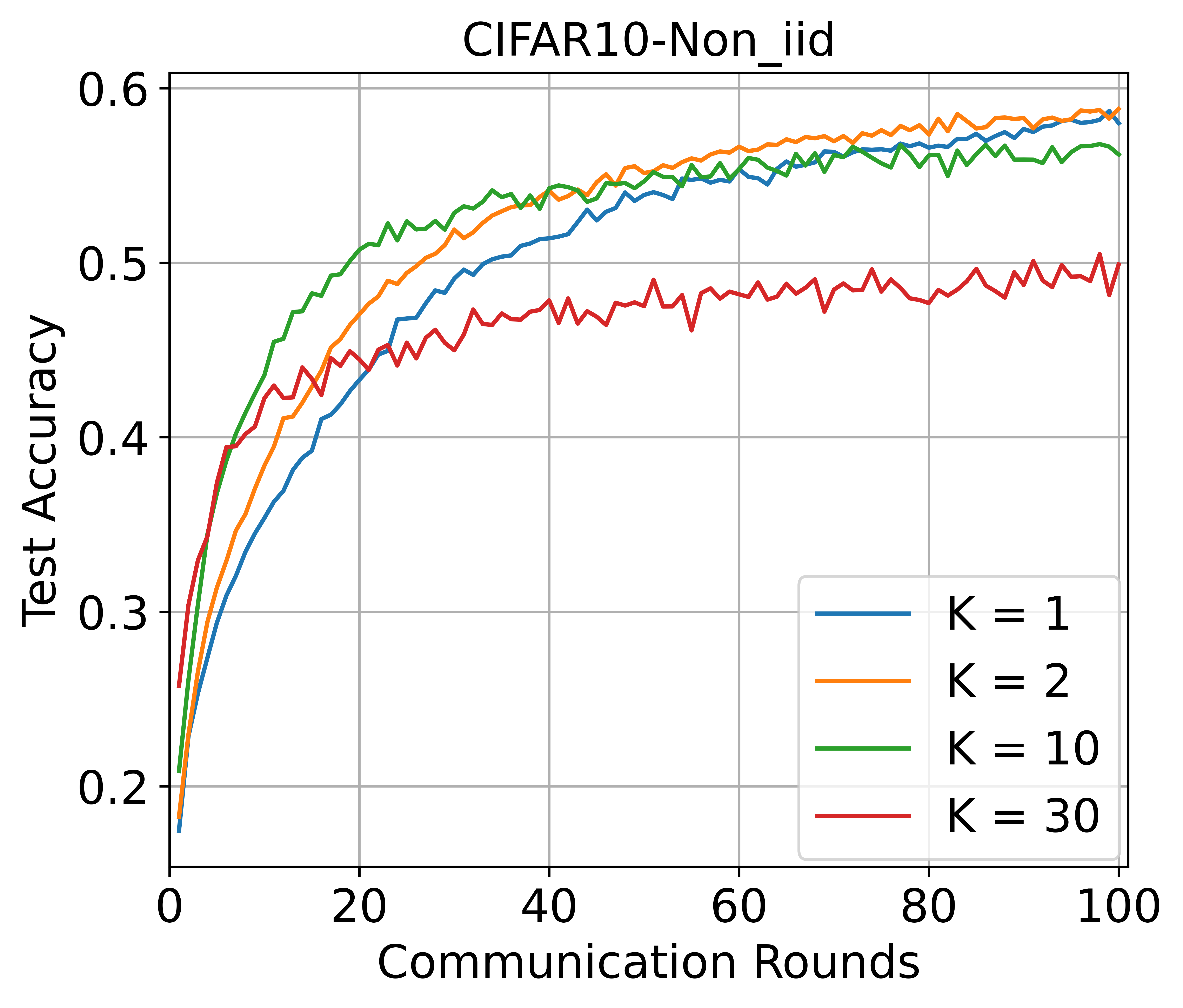}
				\centerline{(b) CIFAR10-Non-IID}
			\end{minipage}
			\vspace{-3mm}
			\caption{Influence of the number of clusters on the training based on the ring topology in the case of heterogeneous resources.}
			\vspace{-6mm}
			\label{mt2}
		\end{figure}
		
		\textbf{Observation 2.} \textit{{In the case of heterogeneous resources across devices, the model accuracy can be effectively improved by organizing the communication between devices in a ring topology, wherein devices are ordered ascendingly or descendingly based on their local training time}. }

		We consider three different ring based communication topologies, wherein devices are ordered randomly, ascendingly or descendingly based on their local training time (i.e., called small-to-large or large-to-small variants). ``small-to-large' means that devices connect sequentially in order of the time required for local training to complete from shortest to longest. Figure~\ref{mt3} shows that the models trained on large-to-small and small-to-large topologies are far more accurate than those trained on random ring topologies. 
		
		Figure~\ref{mt3}  also shows that when the data stored on each device is Non-IID, the test accuracy of the final model is reduced by about 10\% than that in IID data setting. This is mainly due to the problem of catastrophic forgetting~\cite{DBLP:journals/connection/Robins95} when parameters are passed between devices, which refers to the fact that the model may forget the knowledge acquired on earlier devices after training on multiple  following devices. The above problem can be effectively solved by the server periodically gathering the information of all devices, i.e., combining the centralized FL framework..
		
		\textbf{Observation 3.} \textit{{In the case of heterogeneous resources across devices, }clustering the devices with heterogeneous resources into different classes based on the duration of local training can accelerate the training but may lower the model accuracy.} 
		
		\begin{figure}[h]
			\centering
			\includegraphics[width=\linewidth]{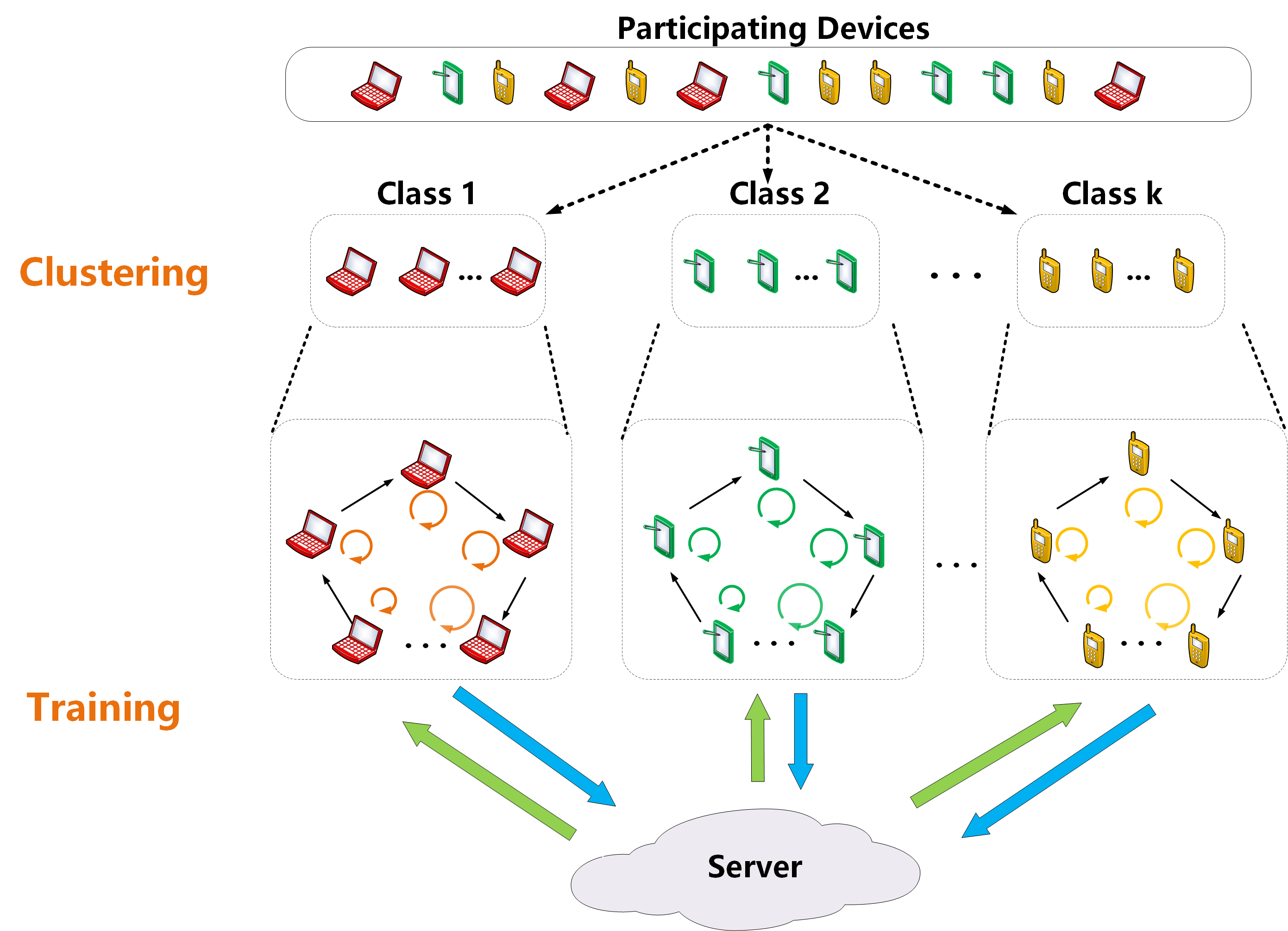}
			\vspace{-6mm}
			\caption{Overview of FedHiSyn.}
			\vspace{-6mm}
			\label{feds}
		\end{figure}
		
		We cluster the 100 devices based on the duration of local training into 1, 2, 10, and 30 categories. Communication between devices in each class follows a ring topology, and each device directly utilizes the model received from its last device for local training. Figure ~\ref{mt2} shows the average accuracy of devices in the most computationally powerful classes. As the number of clusters increases, the number of devices in each category decreases. On the one hand, increasing the number of clusters $K$  allows the device class with strong computing power to complete more communication within the specified time, which speeds up the model training. On the other hand, this leads to less knowledge of data on different devices that the models in each class can learn, affecting the final training accuracy. This is the reason why the accuracy of setting $K$=30 increases the fastest initially but ranks the last finally. In extreme cases, each only one device is a group, which is equivalent to the case where no communication between devices is performed. 
		
		
		%

		\section{FEDHISYN FRAMEWORK}
		FedHiSyn consists of three main components: (1) a centralized server for training the global model; (2)  devices that are clustered into different classes; and (3) a ring topology between devices in a class that ensures the communication direction between devices. 
		
		We now illustrate the training process of FedHiSyn (as depicted in Figure ~\ref{feds} and listed in Alg 1). As shown in Figure ~\ref{feds}  the server completes the device classification and ring topology organization tasks at the beginning of each round. 
		The server clusters devices participating in FL into different classes  based on their response latencies dependent on their computing capacity: $\left\{  class_1, calss_2, ...,class_k \right\} $, where $class_1$ is the fastest class and $class_K$ is the slowest class {(Line 4 in Alg 1)}. The response latency of each device, mainly consisting of its local training time, is recorded by the server.
		
		\begin{figure}[t]
			\vspace{-3mm}
			\begin{algorithm}[H]
				\caption{The training process of FedHiSyn}
				\label{alg:algorithm2}
				
				\begin{algorithmic}[1] 
					\renewcommand{\algorithmicrequire}{\textbf{Input:}}
					\REQUIRE{Number of categories $K$, set of devices $\mathcal{C}$, the time interval at which the server receives parameters $R$;}
					
					\REQUIRE{In server: initialized global weights $ \mathcal{W}_G^0$, time of device $i$ to complete  the local training $t_i$;}
					\REQUIRE{In device $c_i \in \mathcal{C}$: data $\mathcal{D}_i$, buffer $\mathcal{B}_i$.}

					%
					%
					
					%
					
					\STATE $\mathcal{W}_{1,2,\ldots,C} \gets \mathcal{W}_G^0$
					\FOR{iteration $=1,2,\ldots$}
					
					\STATE $\mathcal{S:}$ a random set of devices from $C$  
					\STATE${class}_{1,2,\ldots,K}\gets $\textit{Cluster}($K,t_{1,2,\ldots, \left\|\mathcal{S}\right\|}$)
					\FOR{  $k \in \left\{ 1,2,\ldots,K \right\}$}
					
					\STATE ${Q}_{k} = \left\{{c_i: c_{i+1}}\right\} \gets  $\textit{Small-to-Large-Ring}( $t_{i \in {class}_k}$)
					
					\ENDFOR
					\FOR{$c_i  \in \mathcal{S}$}
					\STATE $\mathcal{B}_i.\text{\textit{clear}()}$ 
					\STATE $\mathcal{B}_i.\textit{push}$($\mathcal{W}_i$)
					\STATE $R_{c_i}  = R$
					\ENDFOR
					
					\WHILE {$R_{c_i}>0$}
					
					\FOR{$c_i  \in \mathcal{S}$ \textbf{in parallel}}
					\STATE $R_{c_i} \gets R_{c_i} - t_i$
					\STATE $Update$($\mathcal{B}_i.back$(), $\mathcal{D}_i$)


					\STATE send ($\mathcal{B}_i.back()$) to $Q_{k}[c_i]$
					\STATE $\mathcal{B}_{Q_{k}[c_i]}.push({\mathcal{B}_i.back()})$
					\ENDFOR
					
					\ENDWHILE	
					
					\STATE $\mathcal{W}_G$ $\gets$  \textit{Aggregate}($\mathcal{W}_{1,2,\ldots,\mathcal{S}}$)
					\STATE $\mathcal{W}_{1,2,\ldots,C} \gets \mathcal{W}_G$ 
					\ENDFOR
					\STATE fuction  $ Update$($\mathcal{W}$, $\mathcal{D}_i$)
					\STATE \qquad $\mathcal{W}_i^{r+1} = \mathcal{W}_G^{r} - \eta\nabla F_i(\mathcal{W})$
					\STATE fuction  $ \textit{Aggregate}$($\mathcal{W}_{1,2,\ldots,\mathcal{S}}$)
					\STATE \qquad $\mathcal{W}_{{G}} = \sum_{i=1}^{\left\|\mathcal{S}\right\|}\frac{1}{\left\|\mathcal{S}\right\|} \mathcal{W}_i$
				\end{algorithmic}
			\end{algorithm}
			\vspace{-10mm}
		\end{figure}
		The server then makes the devices in each class connected to each other in an ascending order of local training time, i.e., in the small-to-large manner. The size of the round arrow in each class, as shown in Figure 5, represents the time it takes for the device to complete the local training. At the end, the device with the longest local training time is connected to the device with the shortest {(Line 5-6 in Alg 1)}. All the devices constitute the ring topology. When a device completes its local training, the trained parameters are sent to the next device following the ring topology {(Line 7-16 in Alg 1)}. At regular intervals, all devices send trained models to the server for model aggregation {(Line 17 in Alg 1)}. 
		
		Specifically, at the beginning of each training round, the server sends each device the initialized parameter $\mathcal{W}_G$. Every $R$ time, the server aggregates the updated  models it receives from all devices to complete a training round. Each device trains with the following steps at each round: (1) The device receives the model from the server and performs local training to update the model; (2) The device sends the updated model to the next device based on the ring topology; (3) After receiving the model sent by another device, the device trains the received model locally to complete the model update. If the device does not receive any model sent by another device, it will continue to train the model that was last trained locally. Steps (2) - (3) repeat until time $T$ is reached, and all devices upload the local models to the server to complete a round of training. 
		
		Under the FedHiSyn framework, devices are always in a working state. This avoids wasting the computing resources of devices as a result of the straggler problem. At the same time, the communication between devices will make the model learn more information of the data stored on diverse devices, thus speeding up the training process and improving the test accuracy of the final model. 
		
		\subsection{Clustering and Communication Topology }
		The equipment set participating in each round of training is recorded as $\mathcal{S}$. At the beginning of each round, the server receives the configuration information sent by the devices participating in this round of training. This information includes the time $t_i$ for Device $i$ to complete the local training and the communication delay $D_{i,j}$ between Device $i$ and Device $j$. The server first clusters devices into $K$ classes based on the time of each device completing local training and then organizes the devices in a ring topology for each class. The clustering algorithm chosen in our setup is the k-means clustering algorithm. The server then sends the organized topological information to each device.  In the ring topology, devices are organized in an ascending order based on the metric shown in  Eq. (5), where device $i+1$ is the next device of Device $i$ in the ring.  
		\begin{equation}
		M_i = t_i+D_{i,i+1},
		\end{equation} 
		Since the communication delay between devices is difficult to estimate precisely, we consider a simpler case where the communication delay between every two consecutive devices is equal, and therefore, Eq. (5) can be simplified to $M_i = t_i$. 
		
		As described in Section 3, the ring-topology communication between devices can significantly improve the efficiency of communication between devices.  At the same time, in order to reduce the impact of resource heterogeneity on the ring topology, we divide devices into different categories according to the duration of the local training, affected by the computing capacity.  The training time of devices in the same category is similar so that the device group with a short average time to complete local training can perform more communication processes. 
		\subsection{Training at Local devices }
		As described in Section 3, using the received model for local training has higher training accuracy than just for aggregation. Therefore, in FedHiSyn, Device $i$ applies an update according to Eq. (6) after receiving model $\mathcal{W}_n$ sent by the server or device $i-1$:
		\begin{equation}
		\mathcal{W}_i = \mathcal{W}_{G/(i-1)} - \eta\nabla F_i(\mathcal{W}_{G/(i-1)}),	
		\end{equation} 
		where $\mathcal{W}_{G/(i-1)}$ is the received model from the server or Device $i-1$, $\mathcal{W}_i$ is the new model after $r$ rounds of communication and training in device $i$, $F_i(. )$ represents the local loss function in Device $i$ and $\eta$ is the learning rate. 
		If device $i$ does not receive any model sent by another device, it applies an update as Eq. (7):
		\begin{equation}
		\mathcal{W}_i^{r+1} = \mathcal{W}_i^r - \eta\nabla F_i(\mathcal{W}). 
		\end{equation}
		The device will repeat the above training process until a predefined time $T$ is reached. 	
		Based on the device training method, the model can be continuously updated and trained by using the local data of different devices.  
		{The training of each model here can be seen as SGD training on the data of multiple devices that the model has traversed. When the model is uploaded to the server, its risk represented of $\mathcal{W}$ on device $i$ as Eq. (8):
			\begin{equation}
			\tilde{F}_i(\mathcal{W}) = \sum_{j \in \Omega } p_j F_j(\mathcal{W}). 
			\end{equation}
			where $\Omega$ is the set of devices that the model $\mathcal{W}$ has been reached,  $j$ is the device in $\Omega$, and $p_j$ is the weight of the $j$-th device such that $p_j  \ge  0$ and $\sum_{j=1}^{\left\|\Omega\right\|} p_j = 1$. $\tilde{F}_i(\mathcal{W})$ is closer to $F(\mathcal{W})$ than ${F_i(\mathcal{W})}$. 	
		}
		Thus, compared with the traditional FL training method, each model can obtain more information of the overall data distribution when it is uploaded to the server for aggregation, thereby improving the accuracy of the training model and accelerating the training speed. 
		\subsection{Weighted Aggregation }
		All models uploaded to the server have been trained by multiple devices and there is no relationship between the weights of each model uploaded to the server and the amount of data on the device that uploaded the model. The traditional method for determining the weights based on the number of samples on each device cannot meet the needs of FedHiSyn very well. In the  stage have been, we assume the weight on each device to be the same so as to avoid what error caused by determining the weight according to the number of labels on the device when the model is uploaded. The aggregation formula is represented as Eq. (9)
		\begin{equation}
		\mathcal{W}_{{G}}^{r+1} = \sum_{i=1}^{\left\|\mathcal{S}\right\|}\frac{1}{\left\|\mathcal{S}\right\|} \mathcal{W}_i^{r+1},
		\end{equation} 
		where $\mathcal{W}_{{G}}^{r+1}$ is the global model in aggregation round $r$, and $\left\|\mathcal{S}\right\|$ is the number of participating devices. 
		
		When the resource heterogeneity of all devices is very high, a class of devices with shorter local training time than another group will have more opportunities for more rounds of communication. In order not to bias the aggregation parameters to these devices with shorter local training time, FedHiSyn can also consider the average time for different types of devices to complete local training as their corresponding weights for parameter aggregation. The aggregation formula is as in Eq. (10):
		\begin{equation}
		\mathcal{W}_{{G}}^{r+1} = \sum_{i=1}^{\left\|\mathcal{S}\right\|}\frac{l_i}{L} \mathcal{W}_i^{r+1},
		\end{equation} 
		where $l_i$ is the average local training time for the class the device $i$ is in, and $L$ is the sum of the average time for all device classes to complete local training. 
		
		Periodic parameter aggregation by the server can integrate the model information of all devices to a certain extent, and accelerate the training speed. At the same time, periodic parameter aggregation by the server can also reduce the catastrophic forgetting problem mentioned in Observation 2 to a certain extent, and has better model accuracy than decentralized training.
		\section{ Convergence Analysis}
		In this section, we show that FedHiSyn converges to the optimal global solution for strongly convex functions on Non-IID data and has a faster convergence rate than FedAvg. We first introduce the  assumptions and definitions as follows.

		Assumptions 5.1, 5.2, 5.3 and 5.4 have been made by the work \cite{DBLP:conf/iclr/LiHYWZ20}. These Assumptions are used to demonstrate the convergence of FedAvg on Non-IID data.
		
		\textbf{Assumption 5.1.}
		$F_1, \cdots, F_N$ are all $L$-smooth:
		for all $v, w \in \mathcal{W}$, 
		$F_i(v)  \leq F_i(w) + (v - w)^T \nabla F_i(w) + \frac{L}{2} \| v - w\|_2^2$.

		\textbf{Assumption 5.2.}
		$F_1, \cdots, F_N$ are all $\mu$-strongly convex:
		for all $v$ and $w$, $F_i(v)  \geq F_i(w) + (v - w)^T \nabla F_i(w) + \frac{\mu }{2} \| v - w\|_2^2$.

		\textbf{Assumption 5.3.}
		Let $\xi_i^r$ be sampled from the $i$-th device's local data uniformly at random.
		The variance of stochastic gradients in each device is bounded: $ \mathbb{E}\left\| \nabla F_i(\mathcal{W}_i^r,\xi_i^r) - \nabla F_i(\mathcal{W}_i^r) \right\|^2 \le \sigma_i^2$  for $i=1,\cdots,C$.

		\textbf{Assumption 5.4.}
		The expected squared norm of stochastic gradients is uniformly bounded, i.e., $ \mathbb{E}\left\| \nabla F_i(\mathcal{W}_i^r,\xi_i^r) \right\|^2  \le G^2$ for all $i=1,\cdots,C$ and $r=0,\cdots, R-1$.

		Based on these four assumptions, the work \cite{DBLP:conf/iclr/LiHYWZ20} proves the convergence of FedAvg on Non-IID data. In FedHiSyn, the parameters of each model are trained on multiple devices, and the device directly uses the received parameters for local training. The overall training process is equivalent to performing SGD on the model sequentially using  the data of multiple devices in different batches. The risk of parameters $\mathcal{W}_i$ uploaded to the server on Device $i$ is changed from $F_i(\mathcal{W})$ to $\tilde{F}_i(\mathcal{W})$. The training gradient of each round is also difficult to represent. $\nabla \tilde{F}_i(\mathcal{W})$ is defined as follows:
		
		\textbf{Definition 5.1.}
		$\nabla \tilde{F}_i(\mathcal{W})$ is introduced to represent the difference between the parameters sent by the server after training by the devices and the initial value, i.e., $\nabla \tilde{F}_i(\mathcal{W}) = \frac{\mathcal{W}_{initial} - \mathcal{W}_{trained}}{\eta} $.
		
		\textbf{Definition 5.2.}
		The training process of parameters under FedHiSyn is $\mathcal{W}_{trained} = \mathcal{W}_{initial} - \eta\nabla \tilde{F}_i(\mathcal{W}$, $\xi)$  and can be seen as all the data on multiple devices, so $\nabla \tilde{F}_i(\mathcal{W}_i^r,\xi_i^r)  = \nabla \tilde{F}_i(\mathcal{W}_i^r)$,  and $\sigma_k^2 = 0 $.
		
		According to Definition 5.1, there are the following lemma 5.1:
		
		\textbf{Lemma 5.1.}
		Let $\left\|\Omega_i\right\|$ be the size of $\Omega_i$ which is the set of devices that  $\mathcal{W}_i$ has been reached. With Definition 5.1 and Assumption 5.4 we have:
		\begin{equation}
		\left\|\nabla \tilde{F}_i(\mathcal{W}_i^r) \right\|^2  \le (\left\|\Omega_i\right\|-1) G^2,
		\end{equation} 
		
		We analyze the case that all the devices participate in the training. Let the {FedAvg} algorithm terminate after $R$ iterations and return $\mathcal{W}^R$ as the solution.
		
		Definition 5.2 is equivalent to  Assumption 5.3, and Lemma 5.1 is equivalent to Assumption 5.4. In FedHiSyn
		we always seem local epoch $E$ as $1$. We consider the simple case where all devices will participate in training. According to the conclusion of work \cite{DBLP:conf/iclr/LiHYWZ20}, with Assumptions 5.1, 5.2, Definition 5.2 and Lemma 5.1 we can get Theorem 5.1 of FedHiSyn.
		
		\textbf{Theorem 5.1.}
		Let  $L, \mu, \sigma_i, G$ be defined therein. Choose $\kappa = \frac{L}{\mu}$, $\gamma = \max\{8\kappa, E\}$ and the learning rate $\eta_t = \frac{2}{\mu (\gamma+r)}$. 
		Then FedHiSyn with {full device participation} satisfies
		\begin{equation}
		\label{eq:bound_K=N}
		\mathbb{E}\left[ \tilde{F}(\mathcal{W}_R)\right] - F^* 
		\: \leq \: 
		\frac{2\kappa}{\gamma + R -1} \left( \frac{12L \Gamma}{\mu} + \frac{\mu \gamma}{2} \mathbb{E}\|\mathcal{W}_0 - \mathcal{W}^*\|^2 \right),
		\end{equation}
		The detailed proof of Theorem 5.1 refers to the works \cite{DBLP:conf/iclr/LiHYWZ20}.
		{In addition, the proof with partial participation can be easily achieved with a few modifications (see details in ~\cite{reddi2021adaptive}).} $F^*$ is the minimum value of $F$. As $R$ increases, the right side of the equation approaches 0, so FedHiSyn can converge to the global optimum.
		
		At the same time, we find that the smaller $\Gamma$ is, the faster the algorithm converges and the lower the communication cost is. Let  $F_i^*$ be the minimum values of  $F_i$.
		We use the term $\Gamma = F^* - \sum_{i=1}^{C} F_i^*$ for quantifying the degree of Non-IID.
		If the data are IID, then $\Gamma$ obviously goes to zero as the number of samples grows.
		If the data are Non-IID, then $\Gamma$ is nonzero, and its magnitude reflects the heterogeneity of the data distribution.
		FedAvg is a special case of FedHiSyn in the case of device resource isomorphism. $\tilde{F}_i$ of FedHiSyn is closer to ${F}$ than $F_i$ of FedAvg, which has been explained in 4.2. Therefore, $\Gamma$ of FedHiSyn  is smaller than that of FedAvg. The convergence speed of FedHiSyn is faster, and the communication cost of the server is lower.

		\section{Experiments}
		In this section, we first introduce the experimental setup and then present the  evaluation results. 
		\subsection{Experimental Setup}
		
		\textbf{Datasets:} We evaluate FedHiSyn using four popular datasets: MNIST, EMNIST-Letter (EMNIST for short), CIFAR10 and CIFAR100 as described below: 
		\begin{itemize}
			\item {MNIST:} It is a dataset that contains a training set of 60,000 examples and a test set of 10,000 examples. Each example is a 28×28 grayscale image associated with a label	from 10 classes. We partition
			the dataset into 100 devices and follow the same Non-IID setting of MNIST as previous experiments ~\cite{DBLP:journals/corr/abs-2102-02079}. 
			\item {EMNIST-Letters:} It is an image dataset that consists of a training set of 124,800 samples and a test set of 20,800 samples. There are a total of 26 classes.  We use the same number of devices and the Non-IID setting as MNIST. 
			\item {CIFAR10:} It dataset consists of 60,000 32 × 32 colour images in 10 classes, with 6000 images per class. There are 50,000 training images and 10,000 test images. We partition the dataset into 100 devices and follow the same Non-IID setting	of CIFAR-10 as ~\cite{DBLP:journals/corr/abs-2102-02079}. 
			\item {CIFAR100:} It has 100 classes. Each class has 600 color images of size 32 × 32, of which 500 are used as training set, and 100 are used as a test set. We use the same number of devices and Non-IID
			setting as CIFAR10.

		\end{itemize}
		The training difficulty of the four datasets is from easy to hard: MNIST, EMNIST, CIFAR10, and CIFAR100. The Non-IID settings that we adopt is Dirichlet($\mathcal{\beta}$): Label distribution on each device follows the Dirichlet distribution, where $\mathcal{\beta}$ is a concentration parameter ($\mathcal{\beta} > 0$).
		\textbf{Models. }	We use fully-connected neural network architectures for MNIST and EMNIST with 2 hidden layers. The number of neurons in the layers is 200 and 100, respectively. For CIFAR-10 and CIFAR-100, we use a CNN model consisting of 2 convolutional layers with 64 5 × 5 filters followed by 2 fully connected layers with 394 and 192 neurons and a softmax	layer. 
		
		\textbf{Baselines. }
		We compare FedHiSyn against six synchronous and asynchronous FL methods:
		\begin{itemize}
			\item  FedAvg ~\cite{DBLP:conf/aistats/McMahanMRHA17}: A baseline synchronous FL method proposed by McMahan \textit{et al.} At each round, a certain ratio of total devices are randomly selected for training, and the server aggregates the weights received from selected devices in an average manner. In an asynchronous setting, FedAvg collects weights from each device at regular intervals, so each device has a different number of rounds of local training. Devices with more computing power are able to do more rounds of local training. 
			\item TFedAvg: A synchronous FL method. After the device finishes training the local model, it waits for other devices to finish local training before uploading the model to the server. The aggregation method of TFedAvg is the same as FedAvg. 
			\item TAFedAvg: An asynchronous FL method, in which each device uploads its local model to the server just after finishing its own training process. The server is responsible for accepting the new models and aggregating them to the original model. 
			\item FedProx ~\cite{DBLP:journals/iotj/RahmanTMT21}: A synchronous FL method that claims to handle stragglers due to system heterogeneity across all devices by applying different local epochs for devices.   FedProx adds a proximal term to the objective on the devices for improving performance and smoothing the training curve. 
			\item FedAT~\cite{DBLP:conf/sc/ChaiC00CR21}: A semi-asynchronous FL method where devices are clustered and assigned to different training layers. Devices within each layer adopt synchronous updates, and different layers directly update asynchronously. 
			\item SCAFFOLD~\cite{DBLP:conf/icml/KarimireddyKMRS20}: A synchronous FL method for Non-IID data, where an additional parameter control variate is added when each device is trained to correct the device-drift presented in FedAvg. 
		\end{itemize}
		\textbf{Metrics. }	We adopt the number of transmitted models between	devices and the server to achieve certain target accuracy as our
		metric to indicate the communication/transmission overheads	and also report the accuracy of the global model achieved in a specified number of rounds.  The overall results are shown in Table 1. Note that the duration of each round in the experiment refers to the time required to complete the local training of the slowest device.  SCAFFOLD communicates the current model and its associated gradient per round, while others communicate only the current model. FedHiSyn, FedAvg, TFedAvg, and FedProx transmit/receive the same number of	models for a fixed number of rounds, whereas SCAFFOLD	costs twice due to the transmission of states. As such number of rounds for SCAFFOLD is one-half of those reported. In each round of training, TAFedAvg and FedAT have more communication between the server and devices, so in this paper, we utilize the average number of transmitted models for all devices in one training round to calculate the final communication volume. The local training performed by different devices differs by a maximum of 10 times. That is to say, before sending an update to the server, a device with strong computing power can communicate with the server 10 times, while a device with weak computing power can only communicate with the server once.  As such number of rounds for FedAT and TAFedAvg is one-fifth of those reported. Along with the transmission costs of each method, we report	the final training accuracy in Table 1.  
		
		\textbf{Hyperparameter setting. } 
		We utilize 100 devices to mimic the large volume of participating edge devices in practical FL. This paper considers three different scenarios where each device has a 100\%, 50\%, and 10\% chance of participating in the training, respectively. The number of epochs for each device to complete local training in one round is randomly distributed in [5, 50]. We set the learning rate as 0. 1, the local mini-batch size as 50, and the local epochs of each device on each round as 5 for TFedAvg, TAFedAvg, and FedAT.  For FedAvg, FedProx and SCAFFOLD, we set the learning rate and local mini-batch size as the same as TFedAvg, but the local epochs for these methods are the maximum achievable training time in a round. 
		
		\begin{figure}[htbp]
			\begin{minipage}[t]{0.5\linewidth}
				\centering
				\includegraphics[width=\textwidth]{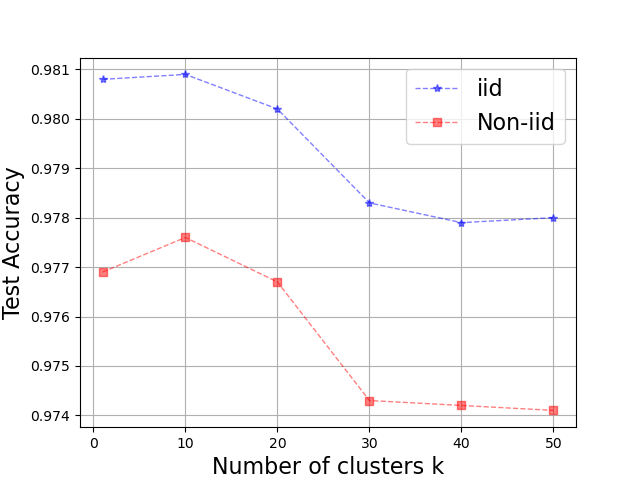}
				\centerline{(a) MNIST}
			\end{minipage}%
			\begin{minipage}[t]{0.5\linewidth}
				\centering
				\includegraphics[width=\textwidth]{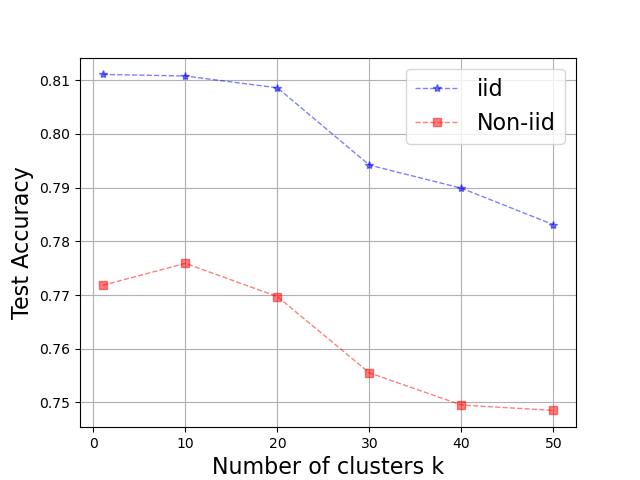}
				\centerline{(b) CIFAR10}
			\end{minipage}
			\vspace{-2mm}
			\caption{Influence of the number K of clustered classes.}
			\vspace{-4mm}
			\label{e1}
		\end{figure}
		
		For FedHiSyn, we set the learning rate, local epoch, and local mini-batch size as the same as FedAT. In the process of device clustering, we choose the K-means clustering algorithm. K is set at 10  when device participation probability is 50\% or 100\% and  is set at 2 when device participation probability is 10\%. Different number of clusters will have an impact on the final training results, which we will discuss later. The clustering options described above are better options. 
		\subsection{Overall results}
		Table ~\ref{tab} shows that FedHiSyn has the optimal communication efficiency in all settings. FedHiSyn can train the model to a predetermined accuracy by transferring fewer parameters to the server. The final FedHiSyn model has the highest test accuracy in the vast majority of settings.  At the same time, we can see that the SCAFFOLD algorithm used to solve the Non-IID data problem still maintains good performance in terms of communication efficiency and model accuracy in the case of heterogeneous resources. 
		
		\textbf{Full vs. Partial Participation. }
		With the Non-IID setting of Dirichlet(0.3), the accuracy of the models trained by FedHiSyn on the four datasets decreases as device engagement decreases from 100\% to 50\% to 10\%. However, compared to FedAvg under the same settings, the accuracy of FedHiSyn is 3.92\%, 4.29\%, and 4.48\% higher on average. We observe similar improvements in performance under most dataset settings. This shows that FedHiSyn has strong robustness to different device participation degrees. When the proportion of devices participating in training is small, the communication between devices can still be used to help train high-precision models.
		
		\textbf{IID vs. Non-IID  Distribution.} 
		Data distributions across devices become more Non-IID as we go from IID, Dirichlet(0.8) to Dirichlet(0.3) settings, which makes the global optimization problem harder. When device participation is 100\%, from the setting of IID, Dirichlet(0.8) to Dirichlet(0.3), the average number of  parameters that FedAvg needs to pass to the server to achieve the target accuracy on the four datasets are 3.4367$\times$, 3.5651$\times$, and 4.1079$\times$ that of FedHiSyn. In the setting of other device engagement, increasing the Non-IID level also results in greater communication savings.
		
		The final training accuracies of FedHiSyn on the IID and Dirichlet (0.3) distributed CIFAR10 datasets are 81.64\% and 79.19\%, respectively, taking on a difference of 2.45\%.  However, The accuracy difference between IID and Non-IID in the experiment of observation 2 is about 10\%. This phenomenon shows that the existence of the server reduces the difference in training accuracy between IID and Non-IID. A plausible explanation is that the server can alleviate the catastrophic forgetting problem in  ring-topology based training.
		
		\begin{figure}[htbp]
			\begin{minipage}[t]{0.5\linewidth}
				\centering
				\includegraphics[width=\textwidth]{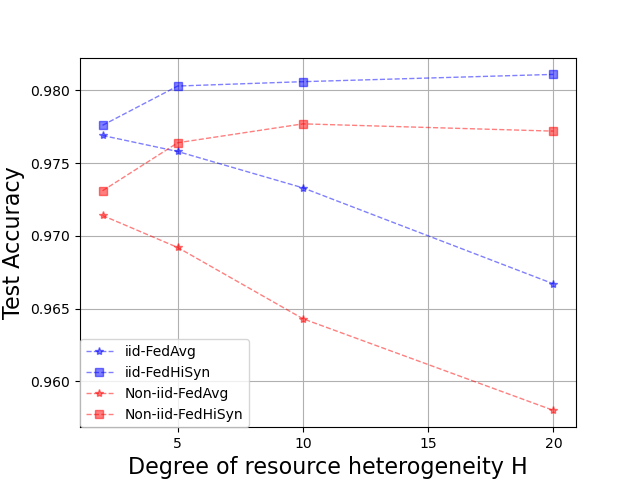}
				\centerline{(a) MNIST}
			\end{minipage}%
			\begin{minipage}[t]{0.5\linewidth}
				\centering
				\includegraphics[width=\textwidth]{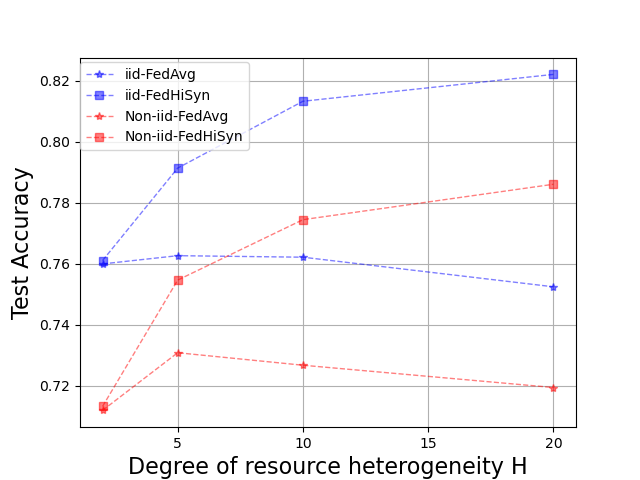}
				\centerline{(b) CIFAR10}
			\end{minipage}
			\vspace{-2mm}
			\caption{Influence of the degree of resource heterogeneity.}
			\vspace{-4mm}
			\label{e2}
		\end{figure}
		
		\textbf{Easy task vs. Difficult task.} 
		
		The more difficult the work FedHiSyn has to do, the more significantly its performance improves over FedAvg. We focus on the case when the device's participation rate is 100\%, and the Non-IID is set to Dirichlet (0.3). Compared with FedAvg, the final training accuracy of FedHiSyn is 1.07\%, 3.13\%, 3.9\%, and 7.58\%  higher regarding the four datasets, respectively.  When the task is more complex, FedHiSyn can still train a model with higher accuracy by using the information transferred between devices.
		\subsection{Effects of the grouping strategy}

		To explore the effect of the number of clustered classes on the model accuracy results, we conduct experiments on the MNIST and CIFAR10 datasets with 50\% devices engagement. As shown in Figure~\ref{e1}, when the number of clusters is set as 1, 10, 20, 30, 40, and 50, respectively, the final model accuracy of FedHiSyn first increases and then decreases, and the highest accuracy is obtained when the number of classes is set at 10. This verifies the motivation that if the number of clustered classes is too small, it is difficult for each class to complete multiple rounds of communication, resulting in a decrease in the efficiency of device communication and a higher accuracy. However, if the number of clustered classes is too large, the number of devices in each category will be too small, and the parameters uploaded to the server will obtain less data information of the devices, which will affect the final training accuracy.
		
		\begin{table*}[]
			\caption{Number of models transmitted, relative to that in one round of FedAvg, to reach a target test accuracy for  participation levels of 100\%, 50\% and 10\%. On the MNIST, EMNIST, CIFAR10, and CIFAR100 datasets, our target test accuracies are 96\%, 86\%, 75\%, and 33\%, respectively. For methods that could not achieve target accuracy within the communication constraint, we append transmission cost with an X sign. The percentage in parentheses represents the final training accuracy. For CIFAR10 and CIFAR100, the number of training rounds is 150, and for MNIST and EMNIST, this is 100. Given the possibility of random device selection during training, the results in the table are averaged based on the results of three independent experiments.}
			
			\centering
			\small
			\resizebox{1\textwidth}{!}{
				\begin{tabular}{|c | c | c c c c c c c |}
					\hline
					Participation & Dataset &  FedHiSyn & FedAvg & FedProx & FedaAT & SCAFFOLD & TAFedAvg  & TFedAvg\\\hline\hline
					\multirow{15}{*}{$\boldsymbol{100\%}$} & \multicolumn{8}{|c|}{\bf{IID}}\\\cline{2-9}
					& \multirow{1}{*}{MNIST}
					& \textbf{5}(98.05\%) &  15(97.21\%) & 15(97.27\%) & 20(97.99\%) & 8(\textbf{98.34\%})	& 25(97.76\%) & 18(97.7\%)   \\\cline{2-9}
					& \multirow{1}{*}{EMNIST}
					& \textbf{4}(90.78\%) &  13(88.55\%) &12(88.55\%) & 15(90.41\%) & 8(\textbf{91.37\%})	& 20(89.77\%) & 13(89.93\%)   \\\cline{2-9}
					& \multirow{1}{*}{CIFAR-10} 
					& \textbf{24}(\textbf{81.64\%}) &  74(77.09\%) & 76(77.35\%) & 120(79.40\%) & 84(79.70\%)	& 225(78.10\%) & 80(78.12\%)   \\\cline{2-9}
					& \multirow{1}{*}{CIFAR-100} 
					& \textbf{29}(\textbf{42.51\%}) &  128(33.59\%) & 118(33.98\%) & 245(37.52\%) & 164(35.33\%)	& 525(34.15\%) & 125(33.97\%)   \\\cline{2-9}
					
					& \multicolumn{8}{|c|}{\bf{ Dirichlet (0.8)}}\\\cline{2-9}
					& \multirow{1}{*}{MNIST}
					& \textbf{6}(97.98\%) &  20(96.96\%) & 19(97.03\%) & 30(97.83\%) & 10(\textbf{98.22\%})	& 35(97.54\%) & 23(97.54\%)   \\\cline{2-9}
					& \multirow{1}{*}{EMNIST}
					& \textbf{5}(90.50\%) &  20(87.63) & 19(87.48\%) & 35(89.32\%) & 10(\textbf{90.95\%})	& 35(89.02\%) & 18(89.50\%)   \\\cline{2-9}
					& \multirow{1}{*}{CIFAR-10} 
					& \textbf{32}(\textbf{80.14\%}) &  95(76.16\%) & 96(76.42\%) & 250(77.80\%) & 92(78.99\%)	& 360(76.63\%) & 105(76.53\%)   \\\cline{2-9}
					& \multirow{1}{*}{CIFAR-100} 
					& \textbf{24}(\textbf{42.85\%}) &  95(34.63\%) & 96(34.63\%) & 210(38.05\%) & 132(36.25\%)	& 430(34.96\%) & 92(35.88\%)   \\\cline{2-9}
					& \multicolumn{8}{|c|}{\bf{ Dirichlet (0.3)}}\\\cline{2-9}
					& \multirow{1}{*}{MNIST}
					& \textbf{8}(\textbf{97.75\%}) &  32(96.68\%) & 29(96.78\%) & 45(97.41\%)  & 12(\textbf{98.15\%}) & 60(97.27\%)	& 29(97.29\%)   \\\cline{2-9}
					& \multirow{1}{*}{EMNIST}
					& \textbf{8}(89.65\%) &  44(86.52\%) & 38(86.83\%) & 65(88.00\%) & 14(\textbf{90.38\%})	& 80(87.67\%) & 25(88.75\%)   \\\cline{2-9}
					& \multirow{1}{*}{CIFAR-10} 
					& \textbf{44}(\textbf{79.19\%}) &  133(75.29\%) & 130(75.36\%) & 440(75.86\%) & 106(78.06\%)	& 610(75.41\%) & 135(75.65\%)   \\\cline{2-9}
					& \multirow{1}{*}{CIFAR-100} 
					& \textbf{22}(\textbf{42.35\%}) &  86(34.77\%) & 89(34.63\%) & 275(36.24\%) & 122(36.70\%)	& 445(34.46\%) & 76(36.09\%)   \\\hline\hline
					
					\multirow{15}{*}{$\boldsymbol{50\%}$} & \multicolumn{8}{|c|}{\bf{IID}}\\\cline{2-9}
					& \multirow{1}{*}{MNIST}
					& \textbf{4}(98.08\%) &  15(97.23\%) & 15(97.31\%) & 45(97.83\%) & 12(\textbf{98.31\%})	& 80(97.38\%) & 18(97.72\%)   \\\cline{2-9}
					& \multirow{1}{*}{EMNIST}
					& \textbf{3}(90.71\%) &  13(88.31\%) & 12(88.46\%) & 30(89.67\%) & 10(\textbf{91.38\%})	& 70(88.39\%) & 13(89.87\%)   \\\cline{2-9}
					& \multirow{1}{*}{CIFAR-10} 
					& \textbf{23}(\textbf{81.70\%}) &  84(76.98\%) & 80(77.08\%) & 275(77.86\%) & 84(79.90\%)	& X(73.59\%) & 85(77.67\%)   \\\cline{2-9}
					& \multirow{1}{*}{CIFAR-100} 
					& \textbf{27}(\textbf{43.65})\% &  145(33.10\%) & 141(33.22\%) & 595(33.13\%) & 180(35.38\%)	& X(27.59\%) & 139(33.46\%)   \\\cline{2-9}
					
					& \multicolumn{8}{|c|}{\bf{ Dirichlet (0.8)}}\\\cline{2-9}
					& \multirow{1}{*}{MNIST}
					& \textbf{6}(97.89\%) &  20(96.96\%) & 20(97.04\%) & 70(97.53\%) & 12(\textbf{98.23\%})	& 125(96.93\%) & 23(97.53\%)   \\\cline{2-9}
					& \multirow{1}{*}{EMNIST}
					& \textbf{5}(90.35\%) &  21(87.58\%) & 20(87.76\%) & 70(88.84\%) & 12(\textbf{91.06\%})	& 145(87.44\%) & 18(89.42\%)   \\\cline{2-9}
					& \multirow{1}{*}{CIFAR-10} 
					& \textbf{30}(\textbf{80.21\%}) &  99(75.92\%) & 104(75.87\%) & 345(76.71\%) & 90(79.24\%)	& X(72.25\%) & 104(76.80\%)   \\\cline{2-9}
					& \multirow{1}{*}{CIFAR-100} 
					& \textbf{23}(\textbf{43.40\%}) &  114(34.04\%) & 118(33.58\%) & 530(33.35\%) & 114(36.13\%)	& X(28.07\%) & 99(35.17\%)   \\\cline{2-9}
					& \multicolumn{8}{|c|}{\bf{ Dirichlet (0.3)}}\\\cline{2-9}
					& \multirow{1}{*}{MNIST}
					& \textbf{8}(97.75\%) &  36(96.59\%) & 35(96.72\%) & 95(97.30\%) & 16(\textbf{97.95\%})	& 225(96.55\%) & 33(97.20\%)   \\\cline{2-9}
					& \multirow{1}{*}{EMNIST}
					& \textbf{8}(89.66\%) &  62(86.34\%) & 44(86.58\%) & 105(87.62\%) & 18(\textbf{90.45\%})	& X(85.66\%) & 26(88.59\%)   \\\cline{2-9}
					& \multirow{1}{*}{CIFAR-10} 
					& \textbf{44}(\textbf{79.07\%}) &  X(74.93\%) & 142(75.02\%) & 595(74.38\%) & 116(77.89\%)	& X(67.97\%) & 139(75.19\%)   \\\cline{2-9}
					& \multirow{1}{*}{CIFAR-100} 
					& \textbf{23}(\textbf{42.63\%}) &  91(34.62\%) & 98(34.54\%) & 490(33.57\%) & 112(36.82\%)	& X(27.92\%) & 84(35.87\%)   \\\hline\hline
					
					\multirow{15}{*}{$\boldsymbol{10\%}$} & \multicolumn{8}{|c|}{\bf{IID}}\\\cline{2-9}
					& \multirow{1}{*}{MNIST}
					& \textbf{5}(97.94\%) &  17(97.13\%) & 17(97.19\%) & 355(96.38\%) & 20(\textbf{98.18\%})	& X(94.43\%) & 19(97.67\%)   \\\cline{2-9}
					& \multirow{1}{*}{EMNIST}
					& \textbf{5}(\textbf{90.27\%}) &  18(87.95\%) & 15(88.15\%) & 240(87.58\%) & 18(90.23\%)	& X(83.38\%) & 15(89.74\%)   \\\cline{2-9}
					& \multirow{1}{*}{CIFAR-10} 
					& \textbf{31}(\textbf{80.32\%}) &  88(76.01\%) & 100(76.03\%) & X(68.94\%) & 108(78.08\%)	& X(57.18\%) & 91(76.91\%)   \\\cline{2-9}
					& \multirow{1}{*}{CIFAR-100} 
					& \textbf{38}(\textbf{41.81\%}) &  X(31.53\%) & X(31.35\%) & X(23.69\%) & X(32.30\%)	& X(16.06\%) & X(32.12\%)   \\\cline{2-9}
					
					& \multicolumn{8}{|c|}{\bf{ Dirichlet (0.8)}}\\\cline{2-9}
					& \multirow{1}{*}{MNIST}
					& \textbf{9}(97.79\%) &  28(96.76\%) & 27(96.86\%) & 460(95.89\%) & 26(\textbf{98.04\%})	& X(92.87\%) & 28(97.36\%)   \\\cline{2-9}
					& \multirow{1}{*}{EMNIST}
					& \textbf{8}(89.72\%) &  38(86.86\%) & 36(87.04\%) & 440(86.28\%) & 28(\textbf{89.67\%})	& X(79.12\%) & 26(88.80\%)   \\\cline{2-9}
					& \multirow{1}{*}{CIFAR-10} 
					& \textbf{46}(\textbf{78.74\%}) &  X(74.03\%) & X(73.92\%) & X(65.55\%) & 146(77.24\%)	& X(51.80\%) & X(74.64\%)   \\\cline{2-9}
					& \multirow{1}{*}{CIFAR-100} 
					& \textbf{36}(\textbf{40.88\%}) &  X(31.68\%) & X(31.91\%) & X(25.02\%) & 254(32.95\%)	& X(16.17\%) & 123(33.45\%)   \\\cline{2-9}
					& \multicolumn{8}{|c|}{\bf{ Dirichlet (0.3)}}\\\cline{2-9}
					& \multirow{1}{*}{MNIST}
					& \textbf{13}(97.15\%) &  48(95.84\%) & 48(96.00\%) & X(94.91\%) & 34(\textbf{97.73\%})	& X(89.58\%) & 36(96.83\%)   \\\cline{2-9}
					& \multirow{1}{*}{EMNIST}
					& \textbf{22}(88.29\%) &  X(84.73\%) & X(85.00\%) & X(84.09\%) & 42(\textbf{88.79\%})	& X(72.93\%) & 38(87.81\%)   \\\cline{2-9}
					& \multirow{1}{*}{CIFAR-10} 
					& \textbf{80}(\textbf{76.98\%}) &  X(72.61\%) & X(72.58\%) & X(62.40\%) & 178(76.57\%)	& X(49.00\%) & X(73.52\%)   \\\cline{2-9}
					& \multirow{1}{*}{CIFAR-100} 
					& \textbf{44}(\textbf{39.66\%}) &  X(30.89\%) & X(31.46\%) & X(24.34\%) & 254(33.04\%)	& X(15.01\%) & 127(33.01\%)   \\\hline
				\end{tabular}
			}
			\label{tab}
		\end{table*}
		
		\subsection{Effects of the degree of heterogeneity}
		
		To explore the effect of resource heterogeneity across devices on the model accuracy, we conduct experiments on the MNIST and CIFAR10 datasets under 50\% device engagement. We use Eq. (10)
		\begin{equation}
		H = l_{max}/l_{min},
		\end{equation}
		to represent the degree of resource heterogeneity of all devices. The greater the H, the greater the degree of resource heterogeneity. As shown in Figure~\ref{e2}, when H is set at 2, 5, 10, and 20, respectively, the accuracies of FedAvg models show an obvious downward trend on the two datasets, while those of FedHiSyn are considerably improved. It can be attributed to that as the degree of resource heterogeneity increases, devices with more computing power have more opportunities to communicate with other devices before sending updates to the server in each round. Considering more data information on different devices is utilized, the accuracy of the final model can be significantly improved.
		
		\section{ Conclusion}
		In this paper, we proposed FedHiSyn, a new framework of FL, to solve the problem of resource and data heterogeneity in FL by utilizing communication between devices. First, devices are grouped into different classes based on the computing capacity and devices within each class are organized in a ring topology. After the clustering, local models are transmitted to and trained on the next devices based on the ring topology before they are uploaded to the server in each iteration. The global partition of devices avoids local communication between devices with high-degree heterogeneous resources, thus relieving the problem of low communication efficiency. By exploiting the local communication between devices, the global model, aggregated from local models, is instilled much knowledge about different data distributions. The proposed framework in this way attenuates the negative effect of the Non-IID problem and saves many communication rounds between the server and devices to reach a target accuracy. Experimental results on various heterogeneous settings on MNIST, EMNIST, CIFAR10, and CIFAR100 datasets have proved that FedHiSyn performs consistently efficiently and effectively in many cases compared to FedAvg, SCAFFOLD, and other baselines.

		
		\bibliographystyle{ACM-Reference-Format}
		\bibliography{fedhisyn}
		

	\end{sloppypar}
\end{document}